\documentclass[conference]{IEEEtran}
\IEEEoverridecommandlockouts
\usepackage{cite}
\usepackage{amsmath,amssymb,amsfonts,mathtools}
\usepackage{algorithmic}
\usepackage{graphicx}
\usepackage{textcomp}
\usepackage{xcolor}
\usepackage{psfrag}
\usepackage{float}
\usepackage{subcaption}

\usepackage[utf8]{inputenc}

\usepackage{multirow}
\usepackage{subcaption}

\newcommand{\norm}[1]{\left\|#1\right\|}

\newcommand{\f}[1]{\boldsymbol{#1}}

\newcommand{\s}[1]{\mathsf{#1}}

\begin{document}
\title{Low PAPR Waveform Design for OFDM Systems Based on Convolutional Autoencoder}

\author{\IEEEauthorblockN{Yara Huleihel}
\IEEEauthorblockA{\textit{Ben Gurion University} \\
\textit{Beer Sheva, Israel}\\
halihal@post.bgu.ac.il} 

% \author{\IEEEauthorblockN{Author 1}
% \IEEEauthorblockA{\textit{Affiliation} \\
% City, Country \\
% Email address}
% %{Address}\\
% %Email}
% \and
% \IEEEauthorblockN{Author 2}
% \IEEEauthorblockA{\textit{Affiliation} \\
% City, Country \\
% Email address}
% \and
% \IEEEauthorblockN{Author 3}
% \IEEEauthorblockA{\textit{Affiliation} \\
% City, Country \\
% Email address}

\and
\IEEEauthorblockN{Eilam Ben-Dror}
\IEEEauthorblockA{\textit{Tel-Aviv Research Center} \\
\textit{Huawei Technologies Co. Ltd.}\\
eilam.ben.dror@huawei.com}
\and
\IEEEauthorblockN{Haim H. Permuter}
\IEEEauthorblockA{\textit{Ben Gurion University} \\
\textit{Beer Sheva, Israel}\\
haimp@bgu.ac.il}
}
\maketitle
\begin{abstract}
This paper introduces the architecture of a convolutional autoencoder ($\mathsf{CAE}$) for the task of peak-to-average power ratio ($\mathsf{PAPR}$) reduction and waveform design, for orthogonal frequency division multiplexing ($\mathsf{OFDM}$) systems. The proposed architecture integrates a $\mathsf{PAPR}$ reduction block and a non-linear high power amplifier ($\mathsf{HPA}$) model. We apply gradual loss learning for multi-objective optimization.
We analyse the model's performance by examining the bit error rate ($\mathsf{BER}$), the $\mathsf{PAPR}$ and the spectral response, and comparing them with common $\mathsf{PAPR}$ reduction algorithms.
\end{abstract}

\vspace{-0.1cm}
\section{Introduction}

Orthogonal frequency division multiplexing ($\mathsf{OFDM}$) has been adopted as a standard technology in various wireless communication systems, such as WiFi, 4G and 5G standards for wireless communications.
Nonetheless, a major drawback of the $\mathsf{OFDM}$ multi-carrier system is its tendency to produce signals with high peak-to-average power ratio ($\mathsf{PAPR}$) in the time-domain, since many subcarrier components are added via a fast Fourier transform ($\mathsf{FFT}$) operation. The contribution of each subcarrier to the total power is dynamic, which makes the total power highly variant.
These high $\mathsf{PAPR}$ signals pass through a non-linear high power amplifier ($\mathsf{HPA}$), which is a major power-consuming analog component, resulting in severe nonlinear signal distortions. Consequently, the resulting signal exhibits spectral regrowth in the form of in-band signal distortions and out-of-band radiation \cite{band_ref}, and the bit error rate ($\mathsf{BER}$) increases. 

The design of $\mathsf{OFDM}$ signals aims to simultaneously achieve high data rate, high spectral efficiency (measured by the adjacent channel power ratio - $\mathsf{ACPR}$) and low computational complexity \cite{ai4b5g, cho, corr, PAPR_5}. This design is highly affected by the nonlinear effects of the $\mathsf{HPA}$. While keeping the $\mathsf{PAPR}$ level low is favorable, there are other design criteria that should be taken into account. Specifically, it is of particular importance to have acceptable signal spectral behavior and $\mathsf{BER}$, which are often referred to as \emph{waveform design}. These criteria often collide, and some trade-offs appear, thus lower $\mathsf{PAPR}$ levels are achieved with higher $\mathsf{BER}$. 

Various $\mathsf{OFDM}$ $\mathsf{PAPR}$ reduction techniques have been proposed in the literature \cite{PAPR_OVER_1,PAPR_OVER_2}. Generally, these techniques can be 
categorized into \emph{model-driven} and \emph{data-driven} techniques. The first category refers to standard approaches in classical communications theory, while the second relies on recent approaches based on machine learning techniques.

\subsection{Classical Approaches (Model Driven)}
$\mathsf{PAPR}$ reduction schemes are roughly classified into three categories: The signal distortion category consists of techniques such as clipping and filtering ($\mathsf{CF}$) \cite{clip1,clip2,clip3,clipSLM}, which limits the peak envelope of the input signal in the time domain to a predetermined value. The multiple signaling probabilistic category includes methods such as selective mapping ($\mathsf{SLM}$) \cite{clipSLM,SLM_PTS}, partial transmit sequence ($\mathsf{PTS}$) \cite{SLM_PTS,PAPR_OVER_1}, ton reservation and ton injection \cite{TI1,PAPR_OVER_1,PAPR_OVER_2}, and constellation shaping \cite{cshape1,cshape2,cshape3}. The main principle of $\mathsf{SLM}$ is to generate several different candidates for each $\mathsf{OFDM}$ block by multiplying the symbols vector with a set of different pseudo-random sequences, and to choose the candidate with the lowest $\mathsf{PAPR}$. In the PTS scheme,  input data are divided into smaller disjoint sub-blocks, which are multiplied by rotating phase factors. The sub-blocks are then added to form the $\mathsf{OFDM}$ symbol for transmission. The objective of PTS is to design an optimal phase factor for a sub-block set that minimizes the $\mathsf{PAPR}$. The coding technique category is presented in \cite{PAPR_OVER_1,PAPR_OVER_2, PAPR_CODING}.

\subsection{Deep Learning Based Schemes (Data Driven)}
In recent years much research has been dedicated to applying deep learning ($\mathsf{DL}$) techniques for designing and optimizing wireless communication networks, e.g. \cite{modelorai, ai4b5g, doublyselective, NN_detector}. Several papers propose $\mathsf{DL}$ methods to handle $\mathsf{PAPR}$ reduction. For example, the authors of \cite{PAPR_Sohn,PAPR_Sohn_Kim}, added a
neural network ($\mathsf{NN}$) 
to reduce the complexity of the active constellation 
scheme, followed by $\mathsf{CF}$.
In \cite{cshape3,AE_DCO} the authors present an autoencoder ($\mathsf{AE}$) solution for $\mathsf{PAPR}$ reduction, while minimizing the $\mathsf{BER}$ degradation. 
The authors in \cite{AE_SLM_DCO,AE_SLM_ACO} proposed a deep $\mathsf{NN}$ combined with $\mathsf{SLM}$ to mitigate the high $\mathsf{PAPR}$ issue of OFDM signal types. They use an $\mathsf{AE}$ structure to represent the constellation mapping and de-mapping of the transmitted symbols.   

\subsection{Main Contributions}
Some of the aforementioned approaches 
suffer from in-band interference, out-of-band distortions and high computational complexity. 
In this paper we aim to handle the $\mathsf{PAPR}$ problem as an integral part of a waveform design objective. In particular, we design a communication system, which simultaneously achieves $\mathsf{PAPR}$ reduction, acceptable spectral behavior of the PA's output and good $\mathsf{BER}$ performance. 
Novelties we introduce include using a $\mathsf{CAE}$ combined with a gradual loss learning technique to handle the multi-objective optimization of the network, and adding the $\mathsf{HPA}$ effect to an integrated end-to-end communication system. 
We demonstrate our algorithm's results on simulated data, and we compare them with classical methods for $\mathsf{PAPR}$ reduction and waveform design, showing competitive results for all three aforementioned objectives. The proposed algorithm allows performance improvement of future wireless communication systems. 

\section{Problem Definition}

In an OFDM system with ${N}$ complex orthogonal subcarriers, the discrete-time
transmitted $\mathsf{OFDM}$ signal is given by the inverse discrete Fourier transform (IDFT),
\begin{align}
    {x}_n = \frac{1}{\sqrt{{N}}}\sum_{k=0}^{{N}-1}{X}_ke^{j\frac{2\pi k}{{L}{N}}kn}, \;\;  0\le n\le {L}{N}-1,\label{eqn:model}
\end{align}
where 
$\{{X}_k\}_{k=0}^{{N}-1}$ are random input symbols modulated by a finite constellation, and ${L}\geq1$ is the over-sampling factor 
(${L}=1$ is the Nyquist sampling rate).
As shown in \cite{PAPR_OVER_1,PAPR_OVER_2}, oversampling by a factor of four results in a good approximation of the continuous-time $\mathsf{PAPR}$ of complex $\mathsf{OFDM}$ signals. 
The $\mathsf{PAPR}$ of the transmitted signal in \eqref{eqn:model} is defined as the ratio between the maximum peak power and the average power of the $\mathsf{OFDM}$ signal, i.e.,
\begin{align}
% \mathsf{PAPR} \triangleq \frac{\smash{\displaystyle\max_{0 \le n \le {LN}-1}} |{x_n}|^2}{\mathbb{E}|{x_n}|^2},
\mathsf{PAPR} \triangleq \frac{\max_{0 \le n \le {LN}-1} |{x_n}|^2}{\mathbb{E}|{x_n}|^2},
\end{align}
%EBD the max range isn't clear.
where $\mathbb{E}\left[\cdot\right]$ denotes the expectation operator.

As $\mathsf{HPA}$ non-linearity causes spectral regrowth, an important assessment for the spectral purity of the system is the $\mathsf{ACPR}$ criterion, which is the ratio between the power of the adjacent channel and the power of the main channel, defined as \cite{acpr}:
\begin{align}
\mathsf{ACPR} \triangleq \frac{\max{\left(\int_{\s{BW}/2}^{3\s{BW}/2} \s{P_{ss}}(f)\;\mathrm{d}f ,\int_{-3\s{BW}/2}^{\s{BW}/2} \s{P_{ss}}(f)\;\mathrm{d}f\right)}}{\int_{-\s{BW}/2}^{\s{BW}/2}\s{P_{ss}}(f)\;\mathrm{d}f},
\end{align}
where $\s{P_{ss}}(\cdot)$ is the power spectral density ($\mathsf{PSD}$) of the signal at the $\mathsf{HPA}$'s output, and $\s{BW}$ is the main channel bandwidth, which is assumed to be equal to the data signal bandwidth.

\begin{figure}[h!]
\centering
\begin{psfrags}
    \psfragscanon
        \psfrag{A}[][][0.8]{$X_k$}
        \psfrag{B}[][][0.9]{$x_n$}
        \psfrag{C}[][][0.8]{PAPR \& SPECTRAL}
        \psfrag{D}[][][0.8]{OPTIMIZATION}
        \psfrag{E}[][][1]{Enc}
        \psfrag{F}[][][0.9]{Filter}
        \psfrag{G}[][][0.9]{${x}_n^\s{F}$}
        \psfrag{H}[][][0.8]{Power}
        \psfrag{I}[][][0.8]{Amplifier}
        \psfrag{J}[][][1]{$\mathrm{G}\left[\cdot\right]$}
        \psfrag{K}[][][0.9]{${x}_n^\s{P}$}
        \psfrag{L}[][][0.8]{AWGN}
        \psfrag{M}[][][0.7]{CHANNEL}
        \psfrag{N}[][][0.9]{$y_n$}
        \psfrag{O}[][][1]{DFT}
        \psfrag{P}[][][0.8]{$Y_k$}
        \psfrag{Q}[][][1]{DEC}
        \psfrag{R}[][][1]{ML}
        \psfrag{S}[][][0.8]{$\hat{{X}}_k$}
        \psfrag{T}[][][1]{IDFT}
        \psfrag{W}[][][0.8]{$w_n$}
    \psfragscanoff
    \includegraphics[scale = 0.7]{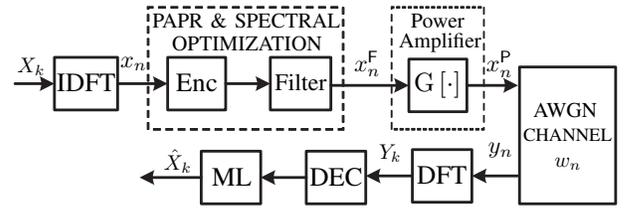}
    \caption{System model diagram.}
    \label{fig:block_model}
\end{psfrags}
\end{figure}

A block diagram of the communication system model is shown in Fig.~\ref{fig:block_model}. Specifically, the encoder and filter blocks mitigate the $\mathsf{PAPR}$ effect and design the waveform to comply with predefined spectral mask requirements. For example, the encoder block can model a clipping operation, while the filter can be a standard band-pass filter ($\mathsf{BPF}$). 
The filtered signal ${x}_n^\s{F}$ is amplified by a non-linear PA.
The amplified signal, ${x}_n^\s{P}=\mathrm{G}({x}_n^\s{F})$, is transmitted through an additive white Gaussian noise ($\mathsf{AWGN}$) channel. 
The channel decoder receives the noisy signal and tries to reconstruct the transmitted signal. Finally, maximum likelihood (ML) is applied for detecting the estimated symbol denoted by $\hat{{X}}_k$. 

The role of the $\mathsf{HPA}$ is to convert the low-level transmission
%EBD Not sure if this should be a separated subsection, as we don't have a B subsection.
signal to a high power signal, capable of driving the antenna at the desired power level.
For achieving maximal power efficiency, the $\mathsf{HPA}$ has to operate close to its saturation region. If the $\mathsf{HPA}$ exceeds the saturation point and enters the non-linear region of operation, the output signal becomes non-linear.
Accordingly, in order to operate the amplifier only in the linear region, we need to make sure that the amplifier operates at a power level that is lower than the saturation point, so that even if the amplifier's input signal increases, the $\mathsf{HPA}$ will not enter the non-linear region. 
This is achieved by down-scaling the input signal by an input back-off (IBO) factor. The drawback of adding the IBO attenuation is that the output power decreases, which makes the $\mathsf{HPA}$ power-inefficient. 

\begin{figure}[h!]
\centering
    % \psfrag{C}[][][1]{}
    \psfrag{M}[][][0.8]{$A_{\mathsf{in}}$}
    \psfrag{Q}[][][0.8]{$A_{\mathsf{out}}$}
    \psfrag{Y}[][][0.6]{$3dB$}
    % \psfrag{G}[][][1]{}
    % \psfrag{B}[][][1]{}
    % \psfrag{A}[][][1]{}
    \includegraphics[width = 8cm]{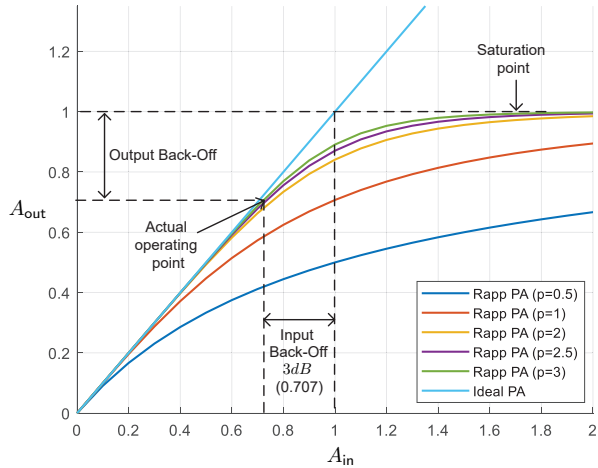}
    \caption{RAPP HPA output versus input signal for different smoothness $p$ values.}
    \label{fig:RAPP_out}
\end{figure}

There are several commonly used models for the non-linearity of an $\mathsf{HPA}$. Here, we will focus on the RAPP model \cite{rapp_1}, which is very accurate for solid-state-power-amplifiers (SSPA), and where only the amplitude is affected. The model's AM/AM conversion is given by 
\begin{align}
    \s{G}({A_{\mathsf{in}}}) = v\cdot{A_{\mathsf{in}}}\cdot\left(1+\left(\frac{v{A}_{\mathsf{in}}}{{A}_0}\right)^{2p}\right)^{-\frac{1}{2p}},
\end{align}
where ${A_{in}}$ is the input amplitude, ${A}_0$ is the limiting output amplitude, $v$ is the small signal gain,
${p}$ is a smoothness parameter controlling the transition from the linear region to the saturation region, 
and $\s{G}({A})$ is the output amplitude. Fig. \ref{fig:RAPP_out} shows RAPP $\mathsf{HPA}$ outputs versus input for several smoothing factor values.
% \begin{figure}[h!]
% \centering
%     \psfrag{A}[][][0.9]{$A_{\mathsf{in}}e^{j{\theta_{in}}}$}
%     \psfrag{B}[][][1]{Envelope's}
%     \psfrag{C}[][][1]{Amplitude}
%     \psfrag{D}[][][0.9]{$A_{\mathsf{in}}$}
%     \psfrag{E}[][][1]{RAPP}
%     \psfrag{F}[][][1]{AM/AM}
%     \psfrag{G}[][][1]{Distortion}
%     \psfrag{H}[][][0.9]{$\;\;A_{out}$}
%     \psfrag{I}[][][0.9]{$     A_{out}e^{j{\theta_{in}}}$}
%     \psfrag{J}[][][1]{${\theta_{in}}$}
%     \psfrag{K}[][][1]{Envelope's}
%     \psfrag{L}[][][1]{Phase}
%     \includegraphics[scale = 0.65]{RAPP_model.eps}
%     \caption{RAPP PA model scheme}
%     \label{fig:RAPP_model}
% \end{figure}

\section{Proposed Waveform Design Structure}
\begin{figure*}[h!]
\centering
    \psfrag{A}[][][0.8]{$X_k$}
    \psfrag{B}[][][0.8]{Zero}
    \psfrag{C}[][][0.8]{pad}
    \psfrag{D}[][][0.8]{IFFT}
    \psfrag{E}[][][0.9]{$x_n$}
    \psfrag{F}[][][0.8]{2}
    \psfrag{G}[][][0.8]{Conv}
    \psfrag{H}[][][0.8]{Layers}
    \psfrag{I}[][][0.8]{Linear}
    \psfrag{J}[][][0.8]{FC}
    \psfrag{K}[][][0.8]{Layer}
    \psfrag{L}[][][0.8]{Power}
    \psfrag{M}[][][0.8]{Norm}
    \psfrag{N}[][][0.8]{Filter}
    \psfrag{O}[][][0.8]{BO}
    \psfrag{P}[][][0.8]{PA}
    \psfrag{Q}[][][0.9]{\;\;$x^F_n$}
    \psfrag{R}[][][0.8]{PAPR}
    \psfrag{S}[][][0.8]{calculation}
    \psfrag{T}[][][0.8]{$ \mathcal{L}_1$}
    \psfrag{U}[][][0.9]{$x^P_n$}
    \psfrag{V}[][][0.8]{AWGN}
    \psfrag{W}[][][0.8]{Channel}
    \psfrag{X}[][][0.8]{$X_k$}
    \psfrag{Y}[][][0.9]{$y_n$}
    \psfrag{Z}[][][0.8]{ACPR}
    \psfrag{a}[][][0.8]{$ \mathcal{L}_2$}
    \psfrag{b}[][][0.8]{loss}
    \psfrag{c}[][][0.8]{$ \mathcal{L}_3$}
    \psfrag{d}[][][0.8]{Reconstruction}
    \psfrag{e}[][][0.8]{Encoder $f(x)$}
    \psfrag{f}[][][0.8]{$\alpha$}
    \psfrag{g}[][][0.8]{Decoder $g(y)$}
    \psfrag{h}[][][0.8]{$\alpha$}
    \psfrag{h}[][][0.8]{$\alpha$}
    \psfrag{v}[][][0.8]{calc}
    \psfrag{k}[][][0.8]{$Y_k$}
    \psfrag{l}[][][0.8]{ML}
    \psfrag{m}[][][0.8]{+ BN}
    \psfrag{n}[][][0.8]{unpad}
    \psfrag{p}[][][0.8]{+ SELU}
    \psfrag{q}[][][0.8]{FFT}
    \psfrag{r}[][][0.8]{PAPR resuction}
    \psfrag{t}[][][1]{Transmitter}
    \psfrag{u}[][][1]{Receiver}
    \psfrag{y}[][][0.8]{Transmitter}
    \psfrag{x}[][][0.8]{$\hat{{X}}_k$}
    \psfrag{w}[][][0.9]{$w_n$}
    \includegraphics[width = 18cm]{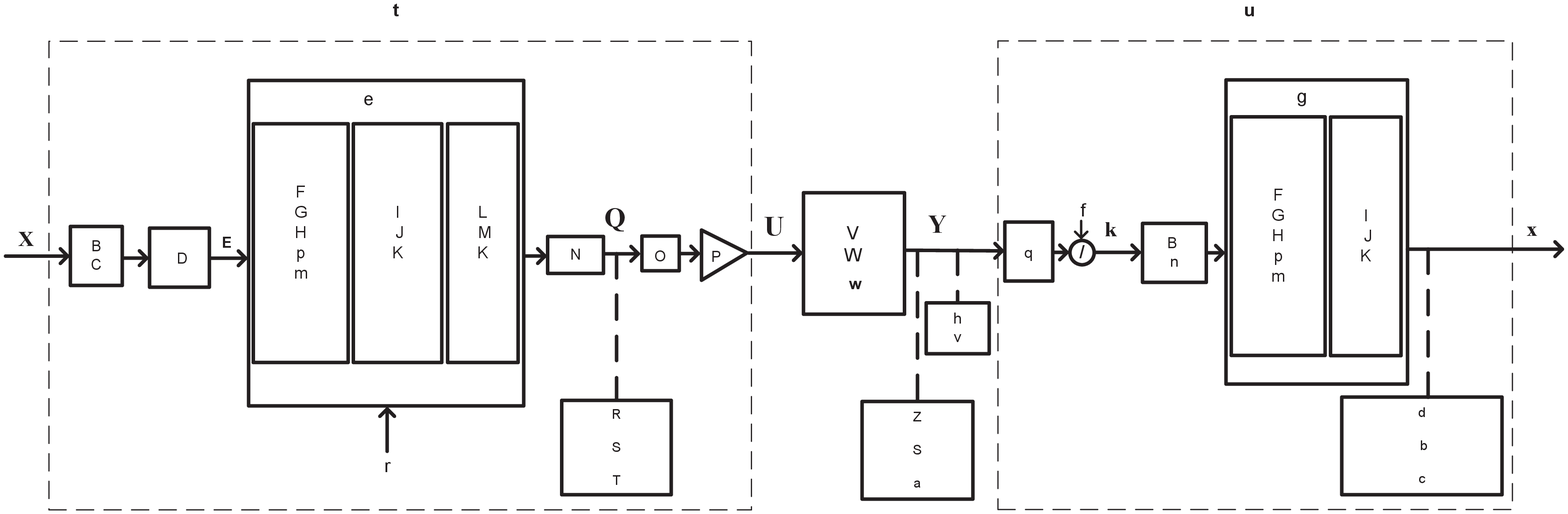}
    \caption{Structure of the proposed conv-AE scheme.}
    \label{fig:NN_scheme}
\end{figure*}

%EBD We propose a $\mathsf{CAE}$ learning system, as depicted in Fig. \ref{fig:NN_scheme}, where the encoder and the decoder of the $\mathsf{CAE}$ consist of 1-dimensional convolutional (Conv1D) layers, and are jointly optimized. 
In this section, we first briefly discuss the $\mathsf{CAE}$ general concept.
The proposed architecture in Fig. \ref{fig:NN_scheme} is then elaborated, including the Bussgang's nonlinearity compensation, followed by a description of the gradual learning process.
  
\subsection{Convolutional Autoencoder (CAE)}
The proposed implementation uses an $\mathsf{AE}$ learning system based on a convolutional neural network (CNN). 
The general structure of an $\mathsf{AE}$ consists of two main blocks: the encoder $f(\f{x})$ and the decoder $g(\f{x})$, where $\f{x}$ is the input data. The $\mathsf{AE}$ is trained to minimize a certain joint loss function, which we denote by $\mathcal{L}(\f{x},g(f(\f{x})))$.
An end-to-end communication system can be interpreted as an $\mathsf{AE}$ in which the encoder and the decoder are part of the transmitter and the receiver respectively, and they can be jointly optimized through an end-to-end learning procedure. 
AEs have been applied in recent years to various wireless communication tasks, such as MIMO detection \cite{MIMO_det}, channel coding \cite{ch_coding} and blind channel equalization \cite{blind_ch}. 

CNNs are widely used for feature extraction and pattern recognition in ML models. Compared with a fully connected ($\mathsf{FC}$) network, a CNN has significantly less connections between adjacent layers, thus less parameters and fewer weights to train, resulting in lower complexity and much faster training. 

% \subsection{Bussgang's Nonlinearity Compensation}
% %EBD Not sure this requires a subsection. Maybe just a paragraph within the proposed CAE architecture subsection, after the AWGN channel.
% To overcome the nonlinearity of the $\mathsf{HPA}$ we compensate the receiver input signal by applying an attenuation factor represented by $\alpha$. The Bussgang's theorem \cite{Bussgang} states that if a zero-mean Gaussian signal passes through a memory-less nonlinear device, then the output-input cross correlation function is proportional to the input autocovariance. Accordingly, the value of $\alpha$ is chosen to minimize the variance of the nonlinear signal distortions. It can be shown that
% \begin{align}
%     \mathsf{\alpha} = \frac{\mathbb{E}\left({x_n}{\overline{s}_n}\right)}{\mathbb{E}\left(|{x_n}|^2\right)},  
% \end{align}
% where ${s_n}$ is the complex output signal of the PA, and ${\overline{s}_n}$ is its complex conjugate.

\subsection{Proposed CAE Architecture}
We propose a $\mathsf{CAE}$ learning system, as depicted in Fig. \ref{fig:NN_scheme}, where the 
input data are the ${N}$ subcarries $\{{X}_k\}_{k=0}^{{N}-1}$ in the frequency domain. Then the signal is zero-padded and converted to the time domain by $\mathsf{FFT}$, outputting $\{{x}_n\}_{n=0}^{{{LN}-1}}$.
These symbols serve as input to the encoder, which acts as a $\mathsf{PAPR}$ reduction block, followed by a filter for optimizing the spectral behavior. 
Both the encoder and the decoder are composed of two convolutional layers, each followed by a non-linear activation function and batch normalization \cite{batch_norm}, and then a fully connected layer.
We have tested several activation functions, including sigmoid, rectified linear unit ($\mathsf{RELU}$), Gaussian error linear unit ($\mathsf{GELU}$), and scaled exponential linear unit ($\mathsf{SELU}$) \cite{SELU}. Empirically, $\mathsf{SELU}$ activation provides the best results for our $\mathsf{CAE}$ scheme.
In addition, the encoder has a power normalization layer \cite{layer_norm}, which 
insures that the transmitted signal meets the power constraints. 

In the transmitter, we use a $\mathsf{BPF}$, whose frequency response is a rectangular window with the same bandwidth as that of ${X}_k$, for reducing the out-of-band radiation. Then, a predefined IBO is applied, and the signal is amplified by the $\mathsf{HPA}$. 
The signal is then transmitted through an $\mathsf{AWGN}$ channel. 

To overcome the nonlinearity of the $\mathsf{HPA}$ we compensate the receiver input signal by applying an attenuation factor represented by $\alpha$. The Bussgang's theorem \cite{Bussgang} states that if a zero-mean Gaussian signal passes through a memory-less nonlinear device, then the output-input cross correlation function is proportional to the input autocovariance. Accordingly, the value of $\alpha$ is chosen to minimize the variance of the nonlinear signal distortions. It can be shown that
\begin{align}
    \mathsf{\alpha} = \frac{\mathbb{E}\left({x_n}{\overline{x}_n^\s{P}}\right)}{\mathbb{E}\left(|{x_n}|^2\right)},  
\end{align}
where ${{x}_n^\s{P}}$ is the complex output signal of the PA, and ${\overline{x}_n^\s{P}}$ is its complex conjugate.

On the receiver side, $\mathsf{FFT}$ converts the signal to the frequency domain, and the signal is divided by $\alpha$ to compensate for the nonlinear distortions.
The zero-unpadding block removes the out of band samples, and finally 
the decoder of the proposed $\mathsf{CAE}$ reconstructs the estimated signal.

\subsection{Training of the CAE Network}
We train a single $\mathsf{CAE}$ model for all tested $\mathsf{SNR}$ values.
We use AdamW optimizer \cite{adamw} that runs back-propagation to optimize the model during training. AdamW is designed such that it improves gradients when using $\text{L}_2$ regularization.

Our loss function is set to optimize three objectives: accurate signal reconstruction (minimal $\mathsf{BER}$), minimal $\mathsf{PAPR}$ and minimal $\mathsf{ACPR}$. These objectives are represented by three loss components $\mathcal{L}_1$, $\mathcal{L}_2$ and $\mathcal{L}_3$, respectively.
That is, 
\begin{align}
    \mathcal{L}(x,\hat{x} ) = \mathcal{L}_1 (x,\hat{x})+\lambda_2 \mathcal{L}_2 (x)+\lambda_3 \mathcal{L}_3 (x),
    \label{eq:loss}
\end{align}
where $\lambda_2$ and $\lambda_3$ are hyper-parameters, which balance the contribution of each loss component to the joint loss function. 

The loss functions we use for optimizing signal reconstruction is the minimum square error (MSE) function with 
$\text{L}_2$ regularization to reduce over-fitting. Denoting by $x$ the input sample (which is also the output target), $\hat{x}$ as the estimated signal, $\Theta$ as the model's weights, and $\lambda_1$ as a hyper-parameter for tuning the $\text{L}_2$ regularization, the loss function is given by
\begin{align}
\mathcal{L}_1 (x,\hat{x}) &= \norm{x-\hat{x}}_2^2+\lambda_1 \norm{\Theta}_2^2.
\end{align}
%EBD Do we need to explain ||2 and ||2^2 ?
For minimizing the $\mathsf{PAPR}$, we calculate it according to the $\mathsf{BPF}$ output, ${{x}_n^\s{F}}$ (cf. Fig. \ref{fig:NN_scheme}), so that 
\begin{align}
    \mathcal{L}_2 (x) & = \mathsf{PAPR}\{{{x}_n^\s{F}}\}.  
\end{align}
The $\mathsf{ACPR}$ loss component is given by
\begin{align}
    \mathcal{L}_3 (x) & = \mathsf{ACPR}\{{{x}_n^\s{P}}\}-\mathsf{ACPR_{req}},  
\end{align}
where, ${{x}_n^\s{P}}$ is the PA's output, and $\mathsf{ACPR_{req}}$ is the required $\mathsf{ACPR}$ value, which is usually dictated by a standard. 
$\mathsf{ACPR_{req}}$ was set according to LTE standard requirements for high spectral purity: $\mathsf{ACPR_{req}}\leq{-45}\mathrm{dB}$ \cite{acpr}. 

We have applied a gradual loss learning technique:
In the first stage, the loss functions consisted only of $\mathcal{L}_1$, so that only the reconstruction loss was optimized. Then, after a predetermined number of epochs, the loss function defined in (\ref{eq:loss}) was used to reduce the $\mathsf{PAPR}$ and improve the spectral behavior.

\section{Results and Insights}
We consider a $\mathsf{SISO}$ $\mathsf{OFDM}$ system with 72 subcarriers and 4-QAM modulation. 4375 batches of 32 samples each were used for training. An oversampling factor $L=4$, and smoothness factor $p=2$ were used.
The structure of the proposed $\mathsf{CAE}$ for the above system is described in Table \ref{CAE_Struct}. 
We compare our $\mathsf{CAE}$ model to a $\mathsf{CF}$ algorithm with clipping ratio of 1.58 dB, and to $\mathsf{SLM}$ with $U=128$ phase sequences.
\vspace{-0.3cm}
\begin{table}[htbp]
    \begin{center}
    \caption{CAE Proposed Structure}
\begin{tabular}{c||c|c|c|c}
     Parameter&value&Kernel&In-channel&Out-channel  \\\hline\hline
     \textbf{Transmitter}  \\\hline
     Input size & 360 \\\hline
     Conv (SELU)& - & 3 & 1 & 13 \\\hline
     Conv (SELU)& - & 3 & 13 & 11 \\\hline
     FC (Linear) size & 360  \\\hline 
     \textbf{Receiver}  \\\hline
     Input size & 72 \\\hline
     Conv (SELU)& - & 3 & 1 & 11 \\\hline
     Conv (SELU)& - & 3 & 11 & 13 \\\hline
     FC (Linear) size & 72  \\\hline
     \textbf{General definitions}  \\\hline
     Conv padding& 2  \\\hline
     Learning rate& 0.001 \\\hline
     epochs num& 160 \\\hline
     Subcarriers number& 72 \\\hline
     $\lambda_2$ & 0.004 \\\hline
     $\lambda_3$ & 0.001 \\\hline
\end{tabular}
\label{CAE_Struct}
\end{center}
\end{table}
\vspace{-0.3cm}
\subsection{BER Analysis}
\begin{figure}[h!]
\centering
\vspace{-0.4cm}
    \psfrag{C}[][][1]{}
    \psfrag{D}[][][1]{$Y=1$}
    \psfrag{E}[][][1]{}
    \psfrag{F}[][][1]{$Y=0$}
    \psfrag{G}[][][1]{$Y=0/1$}
    \psfrag{B}[][][1]{$Q=2$}
    \psfrag{A}[][][1]{$Q=1$}
    \includegraphics[scale = 0.5]{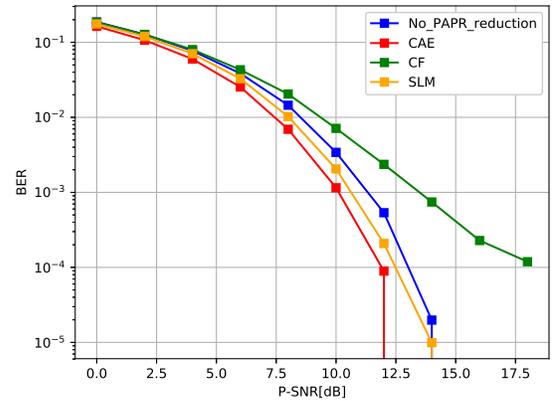}
    \caption{BER vs. $\s{P\_SNR}$ for the considered methods}
    \label{fig:BER_testing}
\end{figure}

Peak Signal to Noise Ratio ($\s{P\_SNR}$) is defined as the ratio between the maximal $\mathsf{HPA}$ power ${A_0}$ and the noise power $\sigma_w^2$, such that
\begin{align}
    \s{P\_SNR} = \frac{A_0^2}{\sigma_w^2}.
\end{align}
% This definition is required, since the actual transmitted power is a function of the transmitter's output back-off ($\mathsf{OBO}$), which evaluates the power efficiency of the system, and defined as the ratio between the maximal $\mathsf{HPA}$ output power ${A_0}$, and the mean transmitted power:
% \begin{align}
%     \mathsf{OBO} = \frac{A_0^2}{\mathbb{E}\left(|{{x}_n^\s{F}}|^2\right)}.
% \end{align}
As shown in Fig. \ref{fig:BER_testing}, the $\mathsf{CAE}$ has better $\mathsf{BER}$ vs. $\s{P\_SNR}$ performance compared to the other examined methods.  

\subsection{CCDF for PAPR Comparison}
\begin{figure}[h!]
\centering
    \psfrag{C}[][][1]{$\mathsf{PAPR_{0}}$}
    \psfrag{P}[][][1]{$\mathsf{PAPR_{0}}[dB]$}
    % \psfrag{E}[][][1]{}
    % \psfrag{F}[][][1]{$Y=0$}
    % \psfrag{G}[][][1]{$Y=0/1$}
    % \psfrag{B}[][][1]{$Q=2$}
    % \psfrag{A}[][][1]{$Q=1$}
    \includegraphics[scale = 0.5]{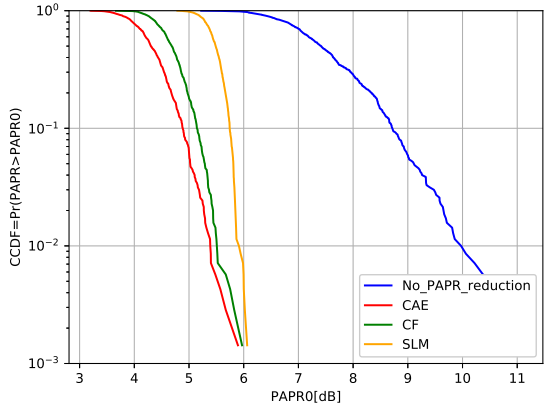}
    \caption{CCDF of PAPR for the considered methods}
    \label{fig:PAPR_testing}
\end{figure}

In order to evaluate the $\mathsf{PAPR}$ performance of different methods, a complementary cumulative distribution function (CCDF) curve is presented in Fig. \ref{fig:PAPR_testing}. The CCDF of the $\mathsf{PAPR}$ denotes the probability that the $\mathsf{PAPR}$ exceeds a certain threshold, i.e. $\mathbb{P}(\mathsf{PAPR} > \mathsf{PAPR_{0}})$.
% \begin{figure}[h!]
% \centering
%     \psfrag{C}[][][1]{$\mathsf{PAPR_{0}}$}
%     \psfrag{P}[][][1]{$\mathsf{PAPR_{0}}[dB]$}
%     % \psfrag{E}[][][1]{}
%     % \psfrag{F}[][][1]{$Y=0$}
%     % \psfrag{G}[][][1]{$Y=0/1$}
%     % \psfrag{B}[][][1]{$Q=2$}
%     % \psfrag{A}[][][1]{$Q=1$}
%     \includegraphics[scale = 0.5]{PAPR_testing_5.eps}
%     \caption{CCDF of PAPR for different PAPR reduction methods}
%     \label{fig:PAPR_testing_5}
% \end{figure}
As can be seen in Fig. \ref{fig:PAPR_testing}, the proposed $\mathsf{CAE}$ achieves better $\mathsf{PAPR}$ reduction compared to the $\mathsf{CF}$ and $\mathsf{SLM}$ methods.

\subsection{Spectrum Analysis}
\begin{figure}[h!]
\centering
    \psfrag{C}[][][1]{}
    \psfrag{D}[][][1]{$Y=1$}
    \psfrag{E}[][][1]{}
    \psfrag{F}[][][1]{$Y=0$}
    \psfrag{G}[][][1]{$Y=0/1$}
    \psfrag{B}[][][1]{$Q=2$}
    \psfrag{A}[][][1]{$Q=1$}
    \includegraphics[scale = 0.5]{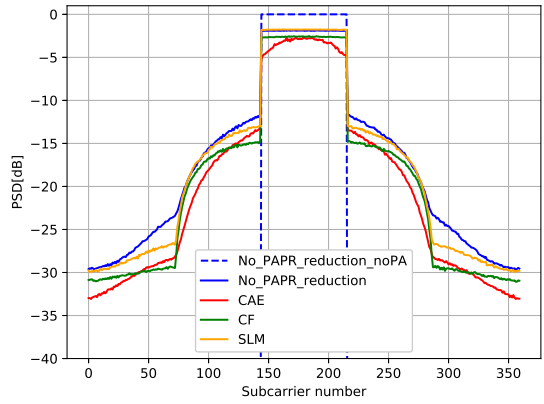}
    \caption{PSD for the considered methods}
    \label{fig:spectral_mask}
\end{figure}

Fig. \ref{fig:spectral_mask} compares the spectral performance  in terms of $\mathsf{PSD}$ of the transmitted signals for all examined methods. The dashed rectangle shows perfect spectral behavior for a linear $\mathsf{HPA}$ with no non-linear components. The proposed $\mathsf{CAE}$ decreases the out-of-band distortions at the expense of lower transmitted power efficiency.
%EBD, which is an acceptable (or valid????) consideration for waveform design systems.   
% \vspace{-0.4cm}
The transmitter's output back-off ($\mathsf{OBO}$), which evaluates the power efficiency of the system, is defined as the ratio between the maximal $\mathsf{HPA}$ output power ${A_0}$, and the mean transmitted power:
\begin{align}
    \mathsf{OBO} = \frac{A_0^2}{\mathbb{E}\left(|{{x}_n^\s{F}}|^2\right)}.
\end{align}
Table \ref{ACPR_OBO} compares the $\mathsf{ACPR}$ and the $\mathsf{OBO}$ of the proposed $\mathsf{CAE}$ to the other methods.
\begin{table}[htbp]
\begin{center}
\caption{ACPR and OBO}
\begin{tabular}{c||c|c|c|c|c}
     Parameter&CAE&FC-AE&CF&SLM&No-reduction  \\\hline\hline
     ACPR[dB]& -28.24 & -25.54 & -29.3 & -27.9  & -26.28\\\hline
     OBO[dB]& 2.5 & 1.58 & 3.34 & 3.5 & 3.7\\
\end{tabular}
 \vspace{-0.4cm}
\label{ACPR_OBO}
\end{center}
\end{table}
As can be seen the $\mathsf{ACPR}$ value of the $\mathsf{CAE}$ is comparable with the considered methods.
It should be noted that for 4-QAM and 72 subcarriers, adding the $\mathsf{ACPR}$ constraint showed only a little improvement. We expect it to have a stronger influence for higher constellations and number of subcarriers.

\begin{figure}[h!]
\centering
\vspace{-0.5cm}
    \psfrag{C}[][][1]{}
    \psfrag{D}[][][1]{$Y=1$}
    \psfrag{E}[][][1]{}
    \psfrag{F}[][][1]{$Y=0$}
    \psfrag{G}[][][1]{$Y=0/1$}
    \psfrag{B}[][][1]{$Q=2$}
    \psfrag{A}[][][1]{$Q=1$}
    \includegraphics[scale = 0.5]{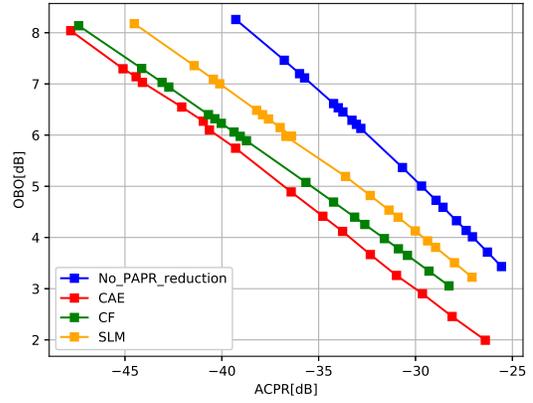}
    \caption{OBO vs. ACPR for the considered methods}
    \label{fig:ACPR_vs_OBO}
\end{figure}

In Fig. \ref{fig:ACPR_vs_OBO} we further compare the $\mathsf{OBO}$ performance for different $\mathsf{ACPR}$ values.  
It can be seen that the CAE system has better power efficiency, while maintaining better BER compared to the other methods.   
\subsection{Autoencoder - FC vs. CNN}
\begin{figure}[h!]
\centering
\vspace{-0.35cm}
    \psfrag{C}[][][1]{}
    \psfrag{D}[][][1]{$Y=1$}
    \psfrag{E}[][][1]{}
    \psfrag{F}[][][1]{$Y=0$}
    \psfrag{G}[][][1]{$Y=0/1$}
    \psfrag{B}[][][1]{$Q=2$}
    \psfrag{A}[][][1]{$Q=1$}
    \includegraphics[trim={0cm  0cm  0cm 0.5cm},clip,scale = 0.5]{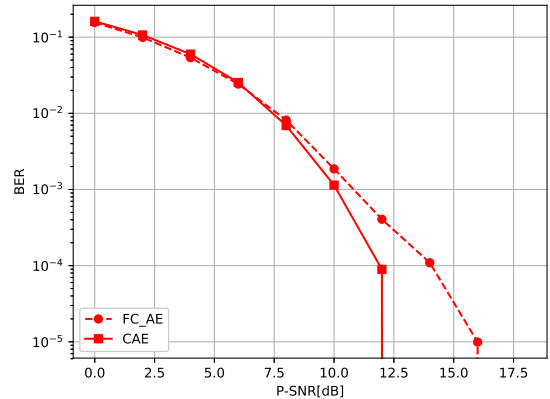}
    \caption{BER vs. $\s{P\_SNR}$ of FC-AE and CAE}
    \label{fig:BER_comp}
\end{figure}

We have investigated various $\mathsf{NN}$ types for the $\mathsf{AE}$, in particular, $\mathsf{FC}$ and $\mathsf{CNN}$. Fig. \ref{fig:BER_comp} compares the $\mathsf{BER}$ performance of two $\mathsf{AE}$ architectures: the proposed $\mathsf{CAE}$, which contains convolutional layers, and a fully connected autoencoder (FC-AE), which contains only $\mathsf{FC}$ layers. 
It can be observed that the $\mathsf{CAE}$ network has better $\mathsf{BER}$ performance compared to the FC-AE. 
As shown in Table \ref{ACPR_OBO}, the $\mathsf{ACPR}$ of the $\mathsf{CAE}$ is better than that of the FC-AE.
Moreover, the $\mathsf{CAE}$ has lower complexity and thus faster training: 
The two convolutional layers have a total of $468$ parameters, while for $\mathsf{FC}$ layers of sizes $2500$ and $3500$, as were used for the FC-AE in Fig. \ref{fig:BER_comp} and Table \ref{ACPR_OBO}, the number of parameters is around $10^7$.

\vspace{-0.4cm}
\subsection{Fixed vs. Gradual Loss Learning}
\begin{figure}[h!]
\vspace{-0.4cm}
\centering
    \psfrag{C}[][][1]{}
    \psfrag{D}[][][1]{$Y=1$}
    \psfrag{E}[][][1]{}
    \psfrag{F}[][][1]{$Y=0$}
    \psfrag{G}[][][1]{$Y=0/1$}
    \psfrag{B}[][][1]{$Q=2$}
    \psfrag{A}[][][1]{$Q=1$}
    \includegraphics[trim={0cm  0cm  0cm 0.5cm},clip,scale = 0.5]{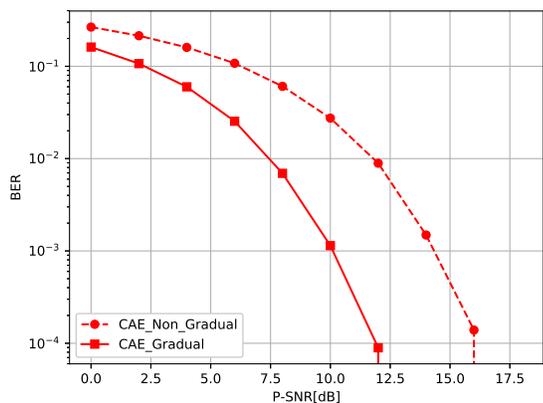}
    \caption{BER vs. $\s{P\_SNR}$ of fixed and gradual loss learning}
    \label{fig:BER_comp_grad}
    \vspace{-0.4cm}
\end{figure}

For showing the benefits of using a gradual loss learning procedure, Fig. \ref{fig:BER_comp_grad} compares its $\mathsf{BER}$ performance to that of a fixed-loss training procedure, where the loss function's weights are fixed for the entire training. It can be observed that the gradual loss learning procedure significantly improves the $\mathsf{BER}$. In addition, improving the $\mathsf{BER}$ while keeping the $\mathsf{PAPR}$ and spectral performance at the desired levels is easier to control when applying the gradual loss learning method than manipulating loss function weights in fixed-loss training.

% \begin{figure}[h!]
% \centering
%     \psfrag{C}[][][1]{}
%     \psfrag{D}[][][1]{$Y=1$}
%     \psfrag{E}[][][1]{}
%     \psfrag{F}[][][1]{$Y=0$}
%     \psfrag{G}[][][1]{$Y=0/1$}
%     \psfrag{B}[][][1]{$Q=2$}
%     \psfrag{A}[][][1]{$Q=1$}
%     \includegraphics[scale = 0.5]{PSD_test_comp.eps}
%     \caption{PSD to compare between FC-AE and $\mathsf{CAE}$}
%     \label{fig:PSD_comp}
% \end{figure}

\section{Conclusions and Future Work}
In this study we have presented a $\mathsf{CAE}$ model for $\mathsf{PAPR}$ reduction and waveform design in an $\mathsf{OFDM}$ system. We have applied a gradual loss learning method to optimize the performance of three objectives: low $\mathsf{BER}$, low $\mathsf{PAPR}$ and adherence to $\mathsf{ACPR}$ spectral requirements. The proposed $\mathsf{CAE}$ was shown to outperform the $\mathsf{CF}$ and the $\mathsf{SLM}$ algorithms. In future work we plan to extend the model to a multiple-input-multiple-output (MIMO) scenario with higher modulation schemes and more complex channel models, aiming to achieve a functional utility for future wireless communication networks.

 \section*{Acknowledgment}
 We would like to express our appreciation and gratitude to Prof. Dov Wulich, Dr. Ilia Yoffe, and Omer Sholev for their professional guidance throughout the work on this paper.

\bibliography{ref}
\bibliographystyle{IEEEtran}

\end{document}